\documentclass[11pt]{amsart}

\usepackage{a4wide}

\usepackage{relsize}

\usepackage{amsfonts,amssymb,amsmath}

\usepackage{hyperref,graphics,graphicx,fancyhdr,color}

\hsize \textwidth
\advance \hsize by -\marginparwidth
\evensidemargin \oddsidemargin
\usepackage{amssymb}
\advance\hoffset by 5mm

\newcommand{\bbR}{{\mathbb R}}

\newcommand{\bbT}{{\mathbb T}}
\newcommand{\eps}{\epsilon}

\newcommand{\om}{\omega}

\newcommand{\ga}{\gamma}
\newcommand{\la}{\lambda}

\newtheorem{df}{Definition}[section]
\newtheorem{corollary}[df]{Corollary}
\newtheorem{lm}[df]{Lemma}

\newtheorem{prop}[df]{Proposition}
\newtheorem{thm}[df]{Theorem}

\newcommand{\We}{W^{\varepsilon}}
\newcommand{\ve}{\varepsilon}

\newcommand{\commentout}[1]{{}}

\newcommand{\cal}{\mathcal}

\makeatother

\begin{document}

\makeatletter \@addtoreset{equation}{section}
\def\theequation{\thesection.\arabic{equation}}

\title{Diffusion limit for a kinetic equation with
  a thermostatted interface}

\author{Giada Basile}
\address{Dipartimento di Matematica\\ Universit\`a di Roma La Sapienza, Roma, Italy}
\email{{\tt basile@mat.uniroma1.it}}

\author{Tomasz Komorowski}
\address{IMPAM, Polish Academy of Sciences\\ Warsaw, Poland}
\email{{\tt tkomorowski@impan.pl}}

\author{Stefano Olla}
\address{Stefano Olla\\
CEREMADE, UMR CNRS\\
Universit\'e Paris-Dauphine, PSL Research University\\
75016 Paris, France}
\email{{\tt olla@ceremade.dauphine.fr}}

\begin{abstract}
We consider  {\em a linear phonon Boltzmann equation} with a
reflecting/transmitting/absorbing 
interface. This equation appears as the Boltzmann-Grad limit
for the energy density function of a harmonic chain of oscillators
with inter-particle stochastic scattering in the presence of a heat bath at
temperature $T$ in contact with one oscillator at the origin.
We prove that under the diffusive scaling the solutions of the phonon equation
tend to the solution $\rho(t,y)$ of a heat equation with the boundary
condition $\rho(t,0)\equiv T$.
\end{abstract}

\thanks{ T.K.  acknowledges the support of the National Science Centre:
NCN grant 2016/23/B/ST1/00492. SO's research is supported by ANR-15-CE40-0020-01 grant LSD. Both T.K. and S.O. were partially supported by the grant 346300 for IMPAN from the Simons Foundation and the matching 2015-2019 Polish MNiSW fund.}


\maketitle

\section{Introduction}

\label{intro}

We consider  {\em a linear phonon Boltzmann equation} in contact with a heat bath at the origin.
This equation describes the evolution, after a proper kinetic limit,
of  the  phonon energy in a chain of harmonic oscillators
with random scattering of velocities, where one oscillator is in contact with a heat bath at temperature $T$.

We denoting  {by}
$W(t,y,k)$ the energy density of phonons at time $t\ge0$,
with respect to their position $y\in\bbR$  and frequency variable
$k\in\bbT$ -  the one dimensional circle, understood as the
 interval $[-1/2,1/2]$  with identified endpoints. The heat bath creates an interface localized at $y=0$.
Outside the interface the density satisfies
\begin{equation}
  \label{eq:8}
 \begin{array}{ll}
 \partial_tW(t,y,k) + \bar\om'(k) \partial_y W(t,y,k) = \ga_0 L W(t,y,k), &
\quad (t,y,k)\in\bbR_+\times \bbR_*\times\bbT_*,
\end{array}
\end{equation}
Here 
$$
\bbR_+:=(0,+\infty), \quad \bbR_*:=\bbR\setminus\{0\},\quad
\bbT_*:=\bbT\setminus\{0\},
$$
The parameter $\ga_0>0$ represents the phonon scattering rate.
The scattering operator $L$, acting only on the $k$-variable,
is given by 
$$
LF(k):= {\displaystyle \int_{\bbT}}R(k,k')
\left[F\left(k'\right) - F\left(k\right)\right]dk',\quad k\in\bbT
$$
for $F$ belongs to $ B_b(\bbT)$ - the set  of bounded measurable,  real valued functions.
Here
\begin{equation}
\label{barom}
\bar\om'(k)=\frac{\om'(k)}{2\pi},\quad k\in\bbT,
\end{equation}
{where  $ \om:\bbT\to[0,+\infty)$ is  {\em the dispersion relation} of the harmonic chain}.

The interface conditions that describe the interaction of a phonon
with a thermostat (placed at $y=0$), at temperature $T>0$, are given as follows:

- the outgoing densities are given in terms of the incoming ones as
\begin{equation}\label{feb1408}
  \begin{split}
W(t,0^+, k)&=p_-(k)W(t,0^+, -k)+p_+(k)W(t,0^-,k)+T\mathfrak g(k), \quad \hbox{ for $0< k\le 1/2$},\\
W(t,0^-, k)&=p_-(k)W(t,0^-,-k) + p_+(k)W(t,0^+, k)+T\mathfrak g(k), \quad \hbox{ for $-1/2< k< 0$}.
\end{split}
\end{equation}
where $p_-,p_+,{\frak g}:\bbT\to(0,1]$ are continuous and
\begin{equation}
\label{012304}
p_+(k)+p_-(k)+  \mathfrak g(k)=1. 
\end{equation}


In other words, $p_-(k)$ and  $p_+(k)$ are the reflection and
transmission coefficients across the interface, respectively. They correspond to the
probabilities of the phonon being reflected, or transmitted, by the interface.
The quantity
$T\mathfrak g(k)$ is the phonon production rate by the thermostat as well as the absorption rate of the
frequency $k$ phonon by the interface. The parameter $T>0$ is the heat bath temperature.
This equation has been obtained in \cite{bos}, without the heat bath,
as the Boltzmann-Grad limit of the energy density 
function for a microscopic model of a heat conductor consisting
of a one dimensional chain of harmonic oscillators, with
inter-particle scattering conserving the energy and  volume.
In the presence of the thermostat, but with no scattering  (the case $\ga_0=0$),
the limit has been proved \cite{kors}.
It is believed that the limit also holds in case of the presence of
scattering, i.e. when $\ga_0>0$.

We are interested in the asymptotics of the solutions
to \eqref{eq:8} under the diffusive scaling, i.e. the limit, as $\eps\to0$, for
$\We(t,y,k) = W(t/\ve^2,y/\ve,k)$, i.e. we consider the equation
\begin{equation}\label{resc:eq}\begin{split}
 &\partial_t \We(t,y,k)
    +\frac 1 {\ve} \; \bar\omega'(k) \partial_y \We(t,y,k)= 
     {\frac{\gamma }{\ve^2}} \int_{\bbT}R(k,k') \left[
       \We\left(t,y,k'\right) - \We\left(t,y,k\right)\right]\;
     dk', \qquad y\neq 0,\\
&\We(0,y,k)=W_0(y,k),
\end{split}\end{equation}  
with the interface conditions \eqref{feb1408}.
Let $R(k) = \int R(k,k') dk'$. 
In our main result, see Theorem \ref{thm011302-19a} below,
we prove that under the assumption 
\begin{equation}
\label{041402-19a}
\int_{\bbT}\frac{\om'(k)^2}{R(k)} dk<+\infty,
\end{equation}
and some other technical hypotheses, formulated in Sections \ref{sec2.2}
and \ref{sec2.3} below,
for any $G\in C_0^\infty(\bbR\times\bbT)$ -compactly supported, smooth
function - we have
\begin{equation}
\label{061502-19}
\lim_{\eps\to0}\int_{\bbR\times\bbT} \We(t,y,k)G(y,k)dydk=\int_{\bbR\times\bbT} \rho(t,y)G(y,k)dydk,
\end{equation}
where
\begin{align}
\label{heat10}
&
\partial_t\rho(t,y)=D\partial_y^2\rho(t,y),\quad (t,y)\in\bbR_+\times\bbR_*,\\
  &
    \rho(t,0^\pm)\equiv T\\
  &
    \rho(0,y)=\rho_0(y):=\int_{\bbT}W_0(y,k)dk. \nonumber
\end{align}
The diffusion coefficient is given by 
\begin{equation}
D = \frac{1}{\gamma} \int \om'(k) (-L)^{-1}\om'(k) \; dk  \label{eq:D}
\end{equation}
that is finite by the assumption \eqref{041402-19a} and the properties of $R(k,k')$ made in section
\ref{sec2.2}.

The result implies that only the absorption and the creation of phonons at the interface
  matter in this diffusive scale. Phonons that are reflected or transmitted will come back to the interface, due to scattering, and eventually get absorbed in a shorter time scale.

The diffusive limit has been considered, without
the presence of
interface, in \cite{b,jko,mmm}. It has been shown there that, if \eqref{041402-19a}
is in force, then the solutions of the initial problem \eqref{resc:eq} converge, as  in \eqref{061502-19},
to $\rho(t,y)$ - the solution of the Cauchy problem for the heat
equation \eqref{heat10}.
When the condition \eqref{041402-19a} is violated a superdiffusive
scaling may be required and the limit could be a fractional
diffusion. This case has been also considered in
\cite{bb,AMP,jko,mmm}.

The case of  the diffusive limit of the solution of a kinetic equation
with an absorbing boundary
  has been considered in e.g.  \cite{LK, BLP, DL, BSS,BBGS}.
The diffusive limit with some other  boundary conditions has
also been discussed in the review paper
\cite{BGS}, see  the references contained
therein. A related result for the radiative transport equation with some
  reflection/transmission condition has been obtained in \cite{BR} for the steady state,
  giving rise to different boundary conditions.
  We are not aware of a similar result in the dynamical case, as considered
  in the present paper.

The result for a fractional diffusive limit with the interface
condition is a subject of the paper \cite{kro}.

\section{Preliminaries and the statement of the main result}

\subsection{Weak solution of the kinetic equation with an interface} 

In what follows we denote $\bbR_-:=(-\infty,0)$. Consider 
$
\widetilde W(t,y,k):=W(t,y,k)-T$. It satisfies the equation
\eqref{eq:8} with the interface given by
\begin{equation}\label{feb1408h}
\widetilde W(t,0^+, k)=p_-(k)\widetilde W(t,0^+, -k)+p_+(k)\widetilde W(t,0^-,k)), \hbox{ for $0< k\le 1/2$},
\end{equation}
and
\begin{equation}\label{feb1410h}
\widetilde W(t,0^-, k)=p_-(k)\widetilde W(t,0^-,-k) + p_+(k)\widetilde W(t,0^+, k), \quad \hbox{ for $-1/2< k< 0$}.
\end{equation}

\begin{df}
\label{df013001-19}
A function
$\widetilde  W:\bar \bbR_+\times\bbR\times \bbT\to\bbR$ is called a (weak)
solution to equation \eqref{eq:8} with the interface conditions
\eqref{feb1408h} and \eqref{feb1410h}, provided that
 it belongs to
$L^2_{\rm loc}(\bbR_+,L^2(\bbR\times\bbT))$, its restrictions
$\widetilde W_\iota$ to  $\bbR_+\times\bbR_{\iota}\times \bbT$, $\iota\in\{-,+\}$
extend to continuous functions on $\bbR_+\times\bar\bbR_{\iota}\times
\bbT$ that satisfy \eqref{feb1408h} and \eqref{feb1410h},
 and
\begin{align}
\label{062504x}
&
0=\int_0^{+\infty}\int_{\bbR\times \bbT}\widetilde   W(t,y,
  k)\left[\partial_t\varphi(t,y,k)+ \; \bar\omega'(k) \partial_y
  \varphi(t,y,k)+\gamma L\varphi(t,y,k)\right]dt dydk\nonumber\\
&+\int_{\bbR\times \bbT}W_0(y,
  k)\varphi(0,y,k)dydk
\end{align}
for any test function $\varphi\in
C_0^\infty(\bar\bbR_+\times\bbR_*\times \bbT)$,
\end{df}

\subsection{Assumption about the dispersion relation and  the
  scattering kernel.} 
\label{sec2.2}

We assume that
$\om(\cdot)$ is even, belongs to $C^\infty(\bbT\setminus\{0\})$,
i.e. it smooth outside $k=0$. Furthermore we assume that $\omega(\cdot)$ is unimodal,
that implies that $k\omega'(k) \ge 0$ for $k\in (-1/2, 1/2)$.

We assume that the scattering kernel is symmetric
\begin{equation}
\label{sym}
R(k,k')=R(k',k),
\end{equation}
positive, except for $0$ frequency, i.e.
\begin{equation}
\label{pos}
R(k,k')>0,\quad (k,k')\in\bbT_*^2
\end{equation}
and the total scattering kernel
\begin{equation}
\label{tot}
R(k):=\int_{\bbT}R(k,k')dk'
\end{equation}
is such that the stochastic kernel
\begin{equation}
\label{stoch}
p(k,k'):=\frac{R(k,k')}{R(k)}\in C^\infty(\bbT^2).
\end{equation}
In addition we assume that \eqref{041402-19a} is in force.

{\bf Example.}  Suppose that
$$
R(k)\sim R_0|\sin(\pi k)|^{\beta},\quad |k|\ll1
$$
for some $\beta\ge 0$ and $R_0>0$
and 
\begin{equation}
\label{A}
\om'(k)\sim 2\om_0' \,{\rm sign}(k)\,|\sin(\pi k)|^{\kappa},\quad |k|\ll 1
\end{equation}
for some $\kappa\ge 0$.
Then \eqref{041402-19a} holds, provided that
\begin{equation}
\label{041402-19}
0\le \beta<1+2\kappa.
\end{equation}

\subsection{About the reflection, transmission and
  absorption coefficients}
\label{sec2.3}

In \cite{kors} the coefficient $p_{\pm}(k)$ and $\mathfrak g(k)$ are obtained from the microscopic dynamics
and depends on the dispersion relation as follows.

Let $\ga>0$ (the thermostat strength) and
\begin{equation}
\label{tg}
\tilde g(\lambda) := \left( 1 + \gamma \int_{\bbT}  \frac{\lambda dk}{\lambda^2 + \omega^2(k)} \right)^{-1}, \quad {\rm Re}\,\la>0.
\end{equation} 
It turns out, see \cite{kors}, that 
\begin{equation}
\label{012410}
|\tilde g(\lambda)|\le 1,\quad \la\in \mathbb C_+:=[\la\in \mathbb C:\, {\rm Re}\,\la>0].
\end{equation} 
The function $\tilde g(\cdot)$ is analytic on $ \mathbb C_+$. 
By Fatou's theorem
we know that
\begin{equation}
\label{nu}
\nu(k) :=\lim_{\eps\to0+}\tilde g(\eps-i\om(k))
\end{equation}
exists a.e. in $\bbT$ and in any $L^p(\bbT)$ sense for
$p\in[1,\infty)$. 
Denote 
\begin{align}
\label{022304}
&\mathfrak g(k) := \frac{\gamma |\nu(k)|^2}{|\bar\om'(k)|^2},
\qquad \mathfrak P(k) := \frac{\gamma \nu(k)}{2|\bar\om'(k)|} \nonumber\\
&
p_+(k):= \left|1 - \mathfrak P(k)\right|^2 \\
&
p_-(k):=\left|\mathfrak P(k)\right|^2.\nonumber
\end{align}
It has been shown  in Section 10 of \cite{kors} that
\begin{equation}
\label{feb1402}
{\rm Re}\,\nu(k)=\left(1+\frac{\ga}{2|\bar\om'(k)|}\right)|\nu(k)|^2.
\end{equation}
This identity implies in particular that
\eqref{012304} is in force.

\subsection{Scaled kinetic equations and the formulation of the main result}

Consider $\We  $ the solution of a rescaled kinetic  equations
\eqref{resc:eq}. Our main result can be stated as follows.
\begin{thm}
\label{thm011302-19a}
Suppose that $W_0(y,k)=T+\widetilde W_0(y,k)$, where $\widetilde W_0\in
L^2(\bbR\times\bbT)$. Under the assumptions made about the scattering kernel
$R(\cdot,\cdot)$ and dispersion relation $\om(\cdot)$,
for any test function $\varphi\in C_0^\infty(\bbR_+\times
\bbR\times\bbT)$ we have
\begin{equation}
\label{011302-19}
\lim_{\eps\to0}\int_0^{+\infty}dt\int_{\bbR\times\bbT}\We(t,y,k)\varphi(t,y,k)dydk=\int_0^{+\infty}dt\int_{\bbR\times\bbT}\rho(t,y)\varphi(t,y,k)dydk,
\end{equation}
where $\rho(\cdot,\cdot)$ is the solution of the heat equation
\begin{align}
\label{heat1}
&
\partial_t\rho(t,y)=D\partial_y^2\rho(t,y),\quad (t,y)\in\bbR_+\times\bbR_*,\nonumber\\
&
\\
&
\rho(0,y)=\rho_0(y):=\int_{\bbT}W_0(y,k)dk,\quad \rho(t,0)\equiv
  T,\,t>0.\nonumber
\end{align}
Here, the coefficient $D>0$ is given by \eqref{def:D1} below.
\end{thm}

Defining $\widetilde \We = \We - T$, one can easily see that it also
satisfies \eqref{resc:eq} with
\begin{equation}\label{feb1408vet}
\widetilde\We(t,0^+, k)=p_-(k) \widetilde\We(t,0^+, -k)+p_+(k) \widetilde\We(t,0^-,k), \hbox{ for $0\le k\le 1/2$},
\end{equation}
and
\begin{equation}\label{feb1410vet}
\widetilde\We(t,0^-, k)=p_-(k) \widetilde\We(t,0^-,-k) + p_+(k) \widetilde\We(t,0^+, k), \quad \hbox{ for $-1/2\le k\le 0$}.
\end{equation}
The initial condition
$
\widetilde\We(0,y,k)=\widetilde W_0(y,k):=W_0(y,k)-T
$ 
belongs to
$L^2(\bbR\times\bbT)$.
This means that it is enough to prove Theorem  \ref{thm011302-19a} for $T=0$.
This proof is presented in Section \ref{sec4}.

\section{Some auxiliaries}

\subsection{Some functional spaces}
Let $H_+^1$ be the Hilbert  obtained as the completion  of the
Schwartz class ${\cal S}(\bbR\times \bbT)$ in the norm
 $$
 \|\varphi\|_{H^1_+}^2:= \|\varphi\|_{L^2(\bbR\times \bbT)}^2+\int_0^{+\infty}\int_{\bbT}|\om'(k)|[\partial_y\varphi(y,k)]^2 dydk
 $$
 Similarly, we introduce $H^1_-$.

Let
 ${\cal H}$ be the Hilbert space obtained as the completion of ${\cal S}(\bbR\times \bbT)$ in the norm
 \begin{equation}
\label{calH}
 \|\varphi\|_{{\cal H}}^2:=  \|\varphi\|_{H^1_-}^2+ \|\varphi\|_{H^1_+}^2.
 \end{equation}
 Let also
 $$
 \|\varphi\|_{{\cal H}_0}^2:= \int_{\bbT}|\om'(k)| \mathfrak g(k)
 \left\{[\varphi(0^+,k)]^2+[\varphi(0^-,k)]^2\right\}dk+ \int_{\bbR\times\bbT^2} [\varphi(y,k')-\varphi(y,k)]^2dy dk dk'
 $$

\subsection{Apriori bounds}

Computing the time derivative we have
\begin{equation}
  \label{eq:14}
  \frac12 \frac{d}{dt} \|\widetilde\We(t)\|_{L^2}^2 = -\frac{\gamma}{2\ve^2} \int_{-\infty}^{\infty} dy \mathcal D(\widetilde\We(t,y, \cdot)) - \frac{1}{2\ve} \int_{\bbT} \; \bar\omega'(k)  \left[\widetilde\We(t,0^-, k)^2-\widetilde\We(t,0^+, k)^2
    \right] dk ,
\end{equation}
with
\begin{equation}
\label{cD}
 \mathcal D(f):=\int_{\bbT^2}R(k,k')[f(k)-f(k')]^2dkdk' .
\end{equation}
Taking into account \eqref{feb1408vet} and \eqref{feb1410vet} we obtain
\begin{align*}
&\int_{\bbT} \; \bar\omega'(k)  \left\{ [\widetilde\We(t,0^-, k)]^2-[\widetilde\We(t,0^+, k)]^2 \right\} dk
\\
&
=\int_{0}^{1/2} \; \bar\omega'(k)  \left\{ [\widetilde\We(t,0^-, k)]^2-\left[p_-(k) \widetilde\We(t,0^+, -k)+p_+(k) \widetilde\We(t,0^-,k)\right]^2 \right\} dk\\
&
+\int_{-1/2}^0 \; \bar\omega'(k)  \left\{ \left[p_-(k) \widetilde\We(t,0^-,-k) + p_+(k) \widetilde\We(t,0^+, k)\right]^2-[\widetilde\We(t,0^+, k)]^2 \right\} dk.
\end{align*}
After straightforward calculations (recall that coefficients $p_\pm(k)
$ are even, while $\bar\omega'(k)$ is odd) we conclude that the right hand
side equals
\begin{align*}
                 \int_{0}^{1/2} \; \bar\omega'(k) & \left\{ \left(\widetilde\We(t,0^-, k)^2+ \widetilde\We(t,0^+, -k)^2\right)
                 \left(1-p_+^2(k)-p_-^2(k)\right)\right. \\
&
\left.-
4p_-(k)p_+(k)\widetilde\We(t,0^+, -k)\widetilde\We(t,0^-,k)\right\} dk.
\end{align*}
Since
$p_+(k)+p_-(k)\le 1$
we have $1-p_+^2(k)-p_-^2(k)\ge 0$. In addition,
\begin{align*}
&
{\rm det}
\left[
\begin{array}{ll}
1-p_+^2(k)-p_-^2(k)&-2p_-(k)p_+(k)\\
&\\
-2p_-(k)p_+(k)&1-p_+^2(k)-p_-^2(k)
\end{array}
\right]
=\left[1-(p_+(k)+p_-(k))^2\right]\left[1-(p_+(k)-p_-(k))^2\right].
\end{align*}
Using \eqref{012304} we conclude that the  quadratic form
$$
(x,y)\mapsto \left(1-p_+^2(k)-p_-^2(k)\right)(x^2+y^2)-4p_-(k)p_+(k)xy
$$
is positive definite as long as $p_+(k)+p_-(k)<1$. The eigenvalues of the form can be determined from the equation
\begin{align*}
&
0
=\left[1-\la-(p_+(k)+p_-(k))^2\right]\left[1-\la-(p_+(k)-p_-(k))^2\right],
\end{align*}
which yields
$$
\la_+:=1-(p_+(k)-p_-(k))^2,\quad \la_-:=1-(p_+(k)+p_-(k))^2
$$
and $\la_+>\la_-$. Note that
$$
2 \mathfrak g(k)\ge \la_-=\mathfrak g(k)\left[1+p_+(k)+p_-(k)\right]\ge \mathfrak g(k).
$$
Equality \eqref{eq:14} allows us to obtain the following apriori bounds
\begin{equation}
  \label{eq:16}
  \begin{split}
    &\|\widetilde\We(t)\|_{L^2(\bbR\times\bbT)}^2 \le  \|\widetilde W_0\|_{L^2(\bbR\times\bbT)}^2,\\
    &\int_0^t ds \int_{\bbR} \mathcal D(\widetilde\We(s,y, \cdot)) dy\le
    \frac{\ve^2}{\gamma}  \|\widetilde W_0\|_{L^2(\bbR\times\bbT)}^2, \\
  &\int_0^t ds  \int_0^{1/2}\bar\om'(k) \mathfrak g(k) dk \; \left(
    \widetilde\We(s,0^-, k)^2 + \widetilde\We(s,0^+,-k)^2\right) \le  \ve  \|\widetilde W_0\|_{L^2(\bbR\times\bbT)}^2.
  \end{split}
\end{equation}
By \eqref{feb1408vet} and \eqref{feb1410vet} we obtain that
\begin{equation}
  \label{eq:2}
  \begin{split}
    \widetilde\We(s,0^+, k)^2 \le \widetilde\We(s,0^-, k)^2 + \widetilde\We(s,0^+, -k)^2, \quad k\in (0,1/2)\\
    \widetilde\We(s,0^-, k)^2 \le \widetilde\We(s,0^-, -k)^2 + \widetilde\We(s,0^+, k)^2, \quad k\in (-1/2,0).
\end{split}
\end{equation}
Then using the unimodality of $\omega(k)$ it follows that
\begin{equation}
  \label{eq:16a}
   \int_0^t ds  \int_{\bbT} dk  |\om'(k)| \mathfrak g(k) \; \left(
     \widetilde\We(s,0^-, k)^2 + \widetilde\We(s,0^+,k)^2\right) \le
   2\ve  \|\widetilde W_0\|_{L^2(\bbR\times\bbT)}^2.
\end{equation}


\subsection{Uniform continuity at $y=0$}

Suppose that $y>0$. 
Let
\begin{equation}
\label{Vep}
V_{\ve}(t,y,k):=\int_0^t\widetilde\We(s,y,k)ds.
\end{equation}
Since $\widetilde\We(s,y, k)$ satisfies  \eqref{resc:eq} we can write
\begin{equation}\label{resc:eq-1}\begin{split}
\ve\left[ \widetilde\We(t,y,k)- \widetilde\We(0,y,k)\right]
    + \; \bar\omega'(k)\partial_yV_{\ve}(t,y,k)= F_\eps(t,y,k),
\end{split}\end{equation} 
with
$$
 F_\eps(t,y,k):={\frac{\gamma }{\ve}} \int_{\bbT} R(k,k')\left[V_{\ve}\left(t,y,k'\right) - V_{\ve}\left(t,y,k\right)\right]\; dk', \qquad y\neq 0.
$$
Hence, using Cauchy-Schwarz inequality, we get
\begin{align*}
&
\|F_\eps(t,\cdot)\|^2_{L^2(\bbR\times\bbT)}=\left({\frac{\gamma
  }{\ve}}\right)^2\int_{\bbR\times\bbT}dydk\left\{\int_0^tds\int_{\bbT}
  R(k,k')\left[\widetilde\We(s,y,k') - \widetilde\We(s,y,k)\right]\;
  dk'\right\}^2\\
&
\le \left({\frac{\gamma}{\ve}}\right)^2\int_{\bbR\times\bbT}dydk\left\{\int_0^tds\int_{\bbT}
  R(k,k')dk'\right\}\left\{\int_0^tds\int_{\bbT}
  R(k,k')\left[\widetilde\We(s,y,k') - \widetilde\We(s,y,k)\right]^2\;
  dk'\right\}\\
&
\le t\|R\|_\infty \left({\frac{\gamma}{\ve}}\right)^2 \int_0^tds\int_{\bbR}\mathcal
   D(\widetilde\We(s,y, \cdot)) dy.
\end{align*}
Using the second estimate of \eqref{eq:16} we conclude that for each $t_0>0$
\begin{equation}
\label{Feps}
\sup_{\eps\in(0,1]}\sup_{t\in[0,t_0]}\|F_\eps(t,\cdot)\|_{L^2(\bbR\times\bbT)} \le
\gamma t_0\|R\|_\infty  \|\widetilde W_0\|_{L^2(\bbR\times\bbT)}^2 <+\infty.
\end{equation}
From \eqref{resc:eq-1} we conclude that
\begin{equation}\label{resc:eq-2}\begin{split}
\partial_yV_{\ve}(t,y,k)= \frac{\tilde F_\eps(t,y,k)}{\om'(k)},\qquad
y, \bar\om'(k)\neq 0,
\end{split}\end{equation} 
where
$$
  \tilde F_\eps(t,y,k)= F_\eps(t,y,k)-\ve\left[ \widetilde\We(t,y,k)- \widetilde\We(0,y,k)\right].
$$
From \eqref{Feps} and  the first estimate of \eqref{eq:16} we conclude
$$
\sup_{\eps\in(0,1]}\sup_{t\in[0,T]}\|\tilde F_\eps(t,\cdot)\|_{L^2(\bbR\times\bbT)}=:\tilde F_*(T)<+\infty.
$$

We have 
\begin{equation}
\label{x-2}
\|V_{\ve}(t,\cdot)\|_{H^1_\pm}\le \|\tilde F_\eps(t,\cdot)\|_{L^2(\bbR\times \bbT)}\quad t\ge0.
\end{equation}
 Since 
$
\dot V_\eps(t)=\widetilde W_\eps(t),
$
from the first estimate of \eqref{eq:16} we conclude that for any $t_0>0$
$$
\sup_{\eps\in(0,1]}\left\|\dot V_{\ve}\right\|_{L^\infty([0,t_0];L^2(\bbR\times\bbT))}<+\infty.
$$
From \eqref{x-2} we get also (cf \eqref{calH})
$$
\sup_{\eps\in(0,1]}\left\|V_{\ve}\right\|_{L^\infty([0,t_0];{\cal H})}<+\infty.
$$
Summarizing, we have shown the following.
\begin{prop}
\label{prop012404}
For any $t_0>0$
\begin{equation}
 \label{012404a}
C(t_0):=\sup_{\eps\in(0,1]}\left(\left\|V_{\ve}\right\|_{L^\infty([0,t_0];{\cal
      H})}+\left\|\dot V_{\ve}\right\|_{L^\infty([0,t_0];L^2(\bbR\times\bbT))}\right)<+\infty
\end{equation}
and 
$$
\lim_{\ve\to0+}\left\|V_{\ve}\right\|_{L^\infty([0,t_0];{\cal H}_0)}=0.
$$
\end{prop}

\bigskip

Denote by $W^{1,\infty}_{0}([0,t_0];L^2(\bbR\times\bbT))$ the
completion of the space of smooth functions   $f:[0,t_0]\to L^2(\bbR\times\bbT)$ satisfying
$f(0)=0$, with respect to the norm 
$$
\|f\|_{W^{1,\infty}_{0}([0,t_0];L^2(\bbR\times\bbT))}:=\sup_{t\in[0,t_0]}\|\dot f\|_{L^2(\bbR\times\bbT)}.
$$
 As a consequence of the above proposition we immediately conclude the following.
\begin{corollary} 
\label{cor012604}
The family $\left(V_{\ve}(\cdot)\right)_{\eps\in(0,1]}$ is bounded in
$
W^{1,\infty}_{0}([0,t_0];L^2(\bbR\times\bbT))\cap L^\infty([0,t_0];{\cal H})
$ for aby $t_0>0$.
Any $\star$-weak limit point $V(\cdot)$ of $V_{\ve}(\cdot)$, as $\ve\to0+$, satisfies 
the following:
\begin{itemize}
\item[1)] $V(t,y,k)\equiv \bar V(t,y):=\mathlarger{\int}_{\bbT}V(t,y,k)dk$ for
  $(t,y,k)\in\bbR_+\times\bbR\times \bbT)$
\item[2)]  the mapping $\bbR_+\times\bbR_\iota\ni (t,y)\mapsto \bar
  V(t,y)$ extends to a mapping from  $C(\bar\bbR_+\times\bar\bbR_\iota)$,
  $\iota\in\{-,+\}$,
\item[3)]  $V(t,0^\pm)=0$ for each $t>0$,
\item[4)] $V(0,y)\equiv 0$, $y\in\bbR$. 
\end{itemize}
\end{corollary}

\section{Proof of Theorem \ref{thm011302-19a}}

\label{sec4}


Thanks to the above estimates, we conclude that the solutions $\widetilde\We(\cdot)$
are $\star$-weakly sequentially compact in $L^\infty\left([0,+\infty);L^2_w(\bbR\times \bbT)\right)$,
where  $L^2_w(\bbR\times \bbT)$ denotes 
$L^2(\bbR\times \bbT)$ with the weak topology.
Suppose that $\bar W(t,y,k)$ is a limiting point for some subsequence $(\widetilde W^{\ve_{n}}(s,y, k))$, 
where $\ve_n\to0$.  For convenience sake we shall denote the
subsequence by $(\widetilde\We(s,y, k))$. Thanks to \eqref{eq:16}
for each $t>0$ we have (cf \eqref{cD})
\begin{equation}
\label{052504}
\lim_{\eps\to0} \int_0^tds\int_{\bbR}{\cal D}\left(\widetilde\We(s,y, \cdot)\right)dy =0
\end{equation}
thus $\bar W(t,y,k)\equiv \rho(t,y)$, for a.e. 
$(t,y,k)\in\bbR_+\times\bbR\times\bbT$.
\begin{lm}
\label{lm031402-19}
Equation
\begin{equation}
\label{cor}
-L X_1=\bar\om'
\end{equation}
has a unique solution such that 
$$
\int_{\bbT}X_1(k)R(k)dk=0
$$
and
\begin{equation}
\label{chi2}
\int_{\bbT}X_1^2(k)R(k)dk<+\infty.
\end{equation}
\end{lm} 
\proof Let $\mu$ be a Borel probability measure on $\bbT$ given by
$$
\mu(dk)=\frac{R(k)}{\bar R}dk,
$$ 
where
$$
\bar R:=\int_{\bbT}R(k)dk.
$$
We can reformulate \eqref{cor} as
\begin{equation}
\label{cor1}
X_1-PX_1=\frac{\bar\om'}{R},
\end{equation}
where, by virtue of \eqref{041402-19a}, the right hand side belongs to $L^2(\mu)$ and $P:L^2(\mu)\to L^2(\mu)$ is a symmetric operator on $L^2(\mu)$ given by
$$
PF(k):=\int_{\bbT}p(k,k')F(k')dk',\quad F\in L^2(\mu).
$$
The operator is a compact contraction
and, since
$$
\int_{\bbT}F(k) (I-P)F(k)F(k)\mu(dk)={\cal D}(F)
$$
we conclude that $1$ is a simple eigenvalue, with the respective
eigenspace spanned on the eigenvector $F_0\equiv 1$. Thus the conclusion
of the lemma follows, as $\bar\om'/R\perp F_0$.
\qed

\bigskip

\begin{prop}
\label{prop021402-19}
For any function $\varphi\in C_0^\infty\left(\bbR_+\times \bbR_*\right)$ we have
\begin{equation}
\label{022404}
\int_0^{+\infty}\int_{\bbR}\rho(t,y)\left[\partial_t\varphi(t,y)+D\partial_{yy}^2 \varphi(t,y)\right]dt dy=0.
\end{equation}
with 
\begin{equation}\label{def:D1}
0<D:=\frac{1}{\ga}\int_{\bbT}  \bar\omega'(k) X_1(k) dk<+\infty.
\end{equation}
\end{prop}
\proof
The claim made in \eqref{def:D1} follows immediately from the fact
that $X_1\not=0$ 
and $D={\cal D}(\chi)$.

To prove \eqref{022404} we apply a version of the perturbed test
function technique. Let $\varphi_\ve(t,y,k)$ be determined by
\begin{equation}
\label{022404a}
\varphi_\ve(t,y,k)=\varphi(t,y) + \ve \chi_1(t,y,k) +  \ve^2 \chi_2(t,y,k), \qquad y\neq 0.
\end{equation}
where $\varphi\in C_0^\infty(\bbR_+\times \bbR_*)$ and $\chi_j(t,y,k)
$, $j=1,2$ are yet  to be determined.

We can write
\begin{align}
\label{062504a}
&
0=\int_0^{+\infty}\int_{\bbR\times \bbT}\partial_t\left[\widetilde\We(t,y, k)\varphi_\ve(t,y,k)\right]dt dydk\nonumber\\
&
=\int_0^{+\infty}\int_{\bbR\times \bbT}\left[\widetilde\We(t,y, k)\partial_t\varphi_\ve(t,y,k)+\partial_t\widetilde\We(t,y, k)\varphi_\ve(t,y,k)\right]dt dydk\nonumber\\
&
=\int_0^{+\infty}\int_{\bbR\times \bbT}\left[\widetilde\We(t,y, k)\partial_t\varphi_\ve(t,y,k)-\frac 1 {\ve} \; \bar\omega'(k) \partial_y \widetilde\We(t,y,k)\varphi_\ve(t,y,k)+\frac{\gamma }{\ve^2}L\widetilde\We(t,y, k)\varphi_\ve(t,y,k)\right]dt dydk
.
\end{align}
Since the support in the $y$ variable of the test function
$\varphi_\eps(t,y,k)$ will turn out to be isolated from $0$,  the integration by parts yield the equation 
\begin{align}
\label{062504}
&
0=\int_0^{+\infty}\int_{\bbR\times \bbT}\widetilde\We(t,y, k)\left[\partial_t\varphi_\ve(t,y,k)+\frac 1 {\ve} \; \bar\omega'(k) \partial_y \varphi_\ve(t,y,k)+\frac{\gamma}{\ve^2}L\varphi_\ve(t,y,k)\right]dt dydk.
\end{align}
Substituting from \eqref{022404a} we obtain that the term in the brackets has the form
\begin{equation}
\label{form}
\frac{1}{\eps}I+I\!I+\eps I\!I\!I_\eps,
\end{equation}
with
\begin{align}
\label{051502-19}
& I:=\gamma L\chi_1+\bar\om'(k)\partial_y\varphi(t,y),\nonumber\\
&I\!I:=\gamma  L\chi_2+\bar\om'(k)\partial_y\chi_1+\partial_t\varphi(t,y),\\
&
I\!I\!I_\eps:=\partial_t\chi_1(t,y,k) +  \ve\partial_t \chi_2(t,y,k)+\bar\om'(k)\partial_y\chi_2. \nonumber
\end{align}
We stipulate that
\begin{equation}
\label{011502-19}
I=0\quad\mbox{and}\quad I\!I=\partial_t\varphi(t,y)+D\partial_{yy}^2\varphi(t,y).
\end{equation}
The first condition yields
\begin{equation}
\label{021502-19}
\chi_1(t,y,k) = -\frac{1}{\ga}\partial_y \varphi(t,y) L^{-1}  \bar\omega'(k) =-\frac{1}{\ga}\partial_y \varphi(t,y)  X_1(k) ,
\end{equation} 
while the second implies that  $\chi_2(t,y,k)$ is the solution of 
\begin{equation}
  \label{eq:3-p}
  \begin{split}
   \ga L\chi_2(t,y,k) =D\partial_{yy}^2\varphi(t,y)- \bar\omega' \partial_y \chi_1(t,y,k) .
  \end{split}
\end{equation}
Substituting from \eqref{021502-19}
we get
\begin{equation}
  \label{eq:3-q}
  \begin{split}
 L\chi_2(t,y,k)   
    =\frac{1}{\gamma}  \left(D +\frac{1}{\ga}\bar\omega' X_1\right) \partial^2_{yy}\varphi(t,y).
 \end{split}
\end{equation}
Since $(D +\bar\omega' X_1/\ga)/R$ belongs to $L^2(\mu)$ and its
orthogonal to constants, we can solve
the equation
$$
LX_2= D +\frac{1}{\ga}\bar\omega' X_1
$$
using the same argument as in  Lemma \ref{lm031402-19}. Then,
$$
\chi_2(t,y,k)   =\frac{1}{\gamma} \partial^2_{yy}\varphi(t,y)X_2(k).
$$
Clearly
$I\!I\!I_\eps=O(1)$. Taking the limit in \eqref{062504} we obtain
\eqref{022404}, that ends the proof of the proposition.\qed

\bigskip

Suppose that $\bar V(t,y)$ is a limiting point for some subsequence
$$
V_{\ve_{n}}(t,y,k)=\int_0^t\widetilde W_{\eps_n}(s,y,k)ds
$$ 
where $\ve_n\to0$, in the sense
  described by Corollary \ref{cor012604}. We can also assume that  the
  respective sequence $(\widetilde W_{\ve}(\cdot))$ $\star$-weakly
  converge to $\rho(t,y)$  in $L^\infty\left([0,+\infty);L^2_w(\bbR\times \bbT)\right)$.

 For convenience sake we
  shall denote the subsequences by $(V_\ve(\cdot))$, $(\widetilde
  W_{\ve}(\cdot))$, respectively. 
 We have
\begin{equation}
\label{051502-19}
 \bar V(t,y)=\int_0^t\rho(s,y)ds,\quad\mbox{ for all }t\ge 0,\,\mbox{ and a.e. }y.
\end{equation}

\begin{prop}
\label{prop011502-19}
For any function $\varphi\in C_0^\infty\left(\bar\bbR_+\times \bbR_*\right)$ we have
\begin{equation}
\label{022404a1}
\int_0^{+\infty}\int_{\bbR}\left\{\bar V(t,y)\left[\varphi(t,y)+D\partial_{yy}^2 \varphi(t,y)\right]-\varphi(t,y)\rho_0(y)\right\}dt dy=0.
\end{equation}
with $D$ given by \eqref{def:D1} and $\rho_0$ defined in  \eqref{heat1}.
\end{prop}
Before showing the proof of the proposition we show how to finish,
with its help, the proof of Theorem \ref{thm011302-19a}.
According to Proposition \ref{prop011502-19}  $\bar V(t,y)$ is a weak solution of the heat equation
$$
\partial_t \bar V(t,y)=D\partial_{yy}^2 \bar V(t,y)+\rho_0(y)
$$
satisfying $\bar V(0,y)\equiv 0$.
According to part 3) of Corollary \ref{cor012604} we
also have $ \bar V(t,0^\pm)\equiv0$, $t\ge0$.
Hence,
$$
\bar V(t,y)=\int_0^t \frac{1}{\sqrt{4\pi Ds}}\int_{\bbR_\pm}\left\{\exp\left\{-\frac{(y-y')^2}{4Ds}\right\}-\exp\left\{-\frac{(y+y')^2}{4Ds}\right\}\right\}\rho_0(y')dy',\quad t,\,\pm y>0.
$$ 
This, combined with \eqref{051502-19}, implies that
\begin{equation}
\rho(t,y)=\frac{1}{\sqrt{4\pi D t}}\int_{\bbR_\pm}\left\{\exp\left\{-\frac{(y-y')^2}{4Dt}\right\}-\exp\left\{-\frac{(y+y')^2}{4Dt}\right\}\right\}\rho_0(y')dy'  ,\quad t,\,\pm y>0 \label{eq:3},
\end{equation}
which satisfies the conclusion of Theorem \ref{thm011302-19a} for $T=0$.
The only thing yet to be shown is the proof of Proposition \ref{prop011502-19}.

\bigskip 

\subsection*{Proof of Proposition \ref{prop011502-19}}
According to \eqref{resc:eq-1} we have
\begin{equation}\label{resc:eq-2}\begin{split}
\partial_tV_{\ve}(t,y,k)
    + \; \frac{1}{\ve}\bar\omega'(k)\partial_yV_{\ve}(t,y,k)= \frac{\ga}{\eps^2}LV_\eps(t,y,k)+ \widetilde\We(0,y,k)
\end{split}\end{equation} 
and obviously $V_\eps(0,y,k)=0$ a.e.
Let $\varphi_\ve(t,y,k)$ be given by \eqref{022404a}.
From \eqref{resc:eq-2} we can write
\begin{align}
\label{062504a}
&
0=\int_0^{+\infty}\int_{\bbR\times \bbT}\partial_t\left[V_\ve(t,y, k)\varphi_\ve(t,y,k)\right]dt dydk\nonumber\\
&
=\int_0^{+\infty}\int_{\bbR\times \bbT}\left\{\left[V_\ve(t,y, k)\partial_t\varphi_\ve(t,y,k)-\frac 1 {\ve} \; \bar\omega'(k) \partial_y V_\ve(t,y,k)\varphi_\ve(t,y,k)\right.\right.\\
&
\left. \left.+\frac{\gamma}{\ve^2}LV_\ve(t,y, k)\varphi_\ve(t,y,k)\right]+\widetilde\We(0,y,k) \varphi_\ve(t,y,k)\right\}dt dydk\nonumber\\
&
=\int_0^{+\infty}\int_{\bbR\times \bbT}\left\{V_\ve(t,y, k)\left[\partial_t\varphi_\ve(t,y,k)+\frac 1 {\ve} \; \bar\omega'(k) \partial_y \varphi_\ve(t,y,k)\right.\right.\nonumber\\
&
\left.\left.+\frac{\gamma }{\ve^2}L\varphi_\ve(t,y,k)\right]+ W_0(y,k)\varphi_\ve(t,y,k)\right\}dt dydk.\nonumber
\end{align}
Substituting from \eqref{022404a} we obtain that the term in the square brackets has the form
\eqref{form},
with
$I$,
$
I\!I
$
and
$
I\!I\!I_\eps
$ as given in \eqref{051502-19}. 
Taking the limit in \eqref{062504a} we obtain \eqref{022404a}.\qed

\end{document}